\documentclass[draft]{agujournal2019}
\usepackage{url} 
\usepackage{lineno}
\usepackage[inline]{trackchanges} 
\usepackage{soul}
\usepackage{amsmath}
\usepackage{amsfonts}
\usepackage{amssymb}
\usepackage{lmodern}
\usepackage{microtype}
\usepackage{listings}
\usepackage{xcolor}
\usepackage{hyperref}
\usepackage{parskip}
\usepackage{booktabs}

\hypersetup{
    colorlinks=true,
    linkcolor=black,
    urlcolor=blue,
    citecolor=black
}

\lstdefinestyle{matlabstyle}{
    basicstyle=\ttfamily\small,
    backgroundcolor=\color{gray!10},
    frame=single,
    framesep=4pt,
    breaklines=true,
    keepspaces=true,
    columns=flexible,
    upquote=true,
}

\draftfalse

\journalname{JGR: Space Physics}

\begin{document}

%
%


\title{Extreme, transient bursts of energy in the auroral ionosphere. I. Predictive radar tracking}


%
%



\authors{Magnus F Ivarsen\affil{1}, Jean-Pierre St-Maurice\affil{1,2}, \color{black} Devin R Huyghebaert\affil{3,1}, \color{black} Yukinaga Miyashita\affil{4,5}, Saif Marei\affil{1}, Jordan Cho\affil{1}, Mahith Madhanakumar\affil{1}, Megan Gillies\affil{6}, Dan Billett\affil{1}, and Glenn C Hussey\affil{1}}

\affiliation{1}{Department of Physics and Engineering Physics, University of Saskatchewan, Saskatoon, Canada}
\affiliation{2}{Department of Physics and Astronomy, University of Western Ontario, London, Canada}
\affiliation{3}{Leibniz Institute of Atmospheric Physics, K\"{u}hlungsborn, Germany}
\affiliation{4}{Center for Heliophysics Research, Korea Astronomy and Space Science Institute, Daejeon, South Korea}
\affiliation{5}{Department of Astronomy and Space Science, Korea University of Science and Technology, Daejeon, South Korea}
\affiliation{6}{Faculty of Science and Technology, Chemistry and Physics, Mount Royal University, Calgary, Canada}

\correspondingauthor{Magnus F Ivarsen}{magnus.fagernes@gmail.com}



\color{black}
\begin{keypoints}
\item Tracking Farley-Buneman threshold exceedances measures ionospheric electric field structure velocities at kilometre and second resolution
\item The four-year speed distribution matches \textit{in-situ} ion drifts where those are reliable, and continues past their noise limit as a power law
\item Field dispersion conditional on threshold exceedance scales with SME from 30 to 60~mV/m; the fastest observed track implies 560 mV/m
\end{keypoints}
\color{black}

%
%

%
%
\begin{abstract}
Three-metre Farley-Buneman irregularities observed by the \textsc{icebear} VHF radar organize into clusters whose apparent motion follows the electric field mapped from the magnetosphere. \color{black} We track these clusters automatically: each is bounded by an $\alpha$-shape at every time step, consecutive frames are associated by an optimal assignment combining shape overlap with a predicted displacement. Births, deaths, splits, and mergers are monitored, and each trajectory is reduced to per-segment velocities by piecewise linear regression. Tracked speeds are validated against in-situ ion drifts measured by DMSP F16 during two conjunctions in May 2021. Across four years of disturbed conditions, the speed distribution of 74,517 tracked clusters agrees with Swarm~A cross-track ion drifts to within a factor of two in probability density for all speeds between 300 and 4000 m/s, and the radar-tracked speed distribution continues as a power law well beyond the noise limit imposed on Swarm by spacecraft attitude jitter. Binning by geomagnetic activity yields a parameterization of the field dispersion conditional on threshold exceedance, \\ $\sigma^2 = 3.68 \times 10^5$~(SME/100 nT)$^{0.353}$~m$^2$ s$^{-2}$, equivalent to 30 to 60~mV/m across the observed activity range, which supplies the amplitude statistics entering the variance term of height-integrated Joule dissipation. During the 10 May 2024 super-storm, on closed field lines equatorward of the dayside cusp, we retrieved an upper-tail sample of this distribution, finding a cluster moving at $11,240\pm660$~m/s and implying a field of approximately 560~mV/m. Together with the unstable fraction of a space weather model's grid volume, our parameterization can in future close the sub-grid contribution to the storm-time heating budget in the auroral ionosphere. \color{black}
\end{abstract}

\section*{Plain Language Summary}

Earth's aurorae, or northern and southern lights, mark the location of electric currents that flow in the upper atmosphere when the solar wind pushes hard against Earth's extended magnetic field. The strongest of these electrical currents --- the auroral electrojets, at roughly 105~km above Earth's surface --- are driven by electric fields that are difficult to measure: satellites pass through them too quickly to map their structure, and ground-based radars typically return difficult-to-interpret plasma motions. \color{black} We present an automated method that finds compact clusters of radar echoes by density clustering, then follows each cluster from frame to frame. The \color{black} clusters correspond directly to plasma turbulence excited by the strong electric fields. The method is made possible by the advanced scientific imaging radar \textsc{icebear}, located in Saskatchewan, Canada. Each tracked radar echo cluster is automatically tracked across consecutive frames; its motion is then converted into a local electric field measurement, validated against four years of polar-orbiting satellite measurements (2020--2024). During the May 2024 geomagnetic super-storm, which was the most powerful storm to occur in two decades, the method detected a transient (fast-appearing and fast-disappearing) event in which several electric structures travelled at over 10{,}000~meters per second, faster than any value previously documented in this region of space. 

\vspace{10pt}

%
%

%


%
%
%
%

\section{Introduction}

In Earth's upper atmosphere, at the lower edge of the partially ionized layers of the ionosphere, intense, small-scale plasma turbulence is observed during \color{black} disturbed conditions, \color{black} in the form of waves that \color{black} grow, steepen, and saturate \color{black} inside the electrical currents belonging to the auroral electrojets. Compared to the large scales of the aurorae and the magnetosphere-ionosphere (MI) coupling, these turbulent waves, a few meters in size, are \textit{microscopic}, and the highly structured, turbulent, plasma that is created around aurorae \color{black} disturbs the propagation of radio signals across a wide range of wavelengths, from HF through L-band \color{black} \cite{bahcivanObservationsColocatedOptical2006,song_investigation_2025}.

Electrojet turbulence, created by Farley-Buneman \citep[FB, ][]{farleyPlasmaInstabilityResulting1963,bunemanExcitationFieldAligned1963} waves, is excited by the externally applied electric field \cite{ivarsen_characteristic_2025}, a field that produces a relative velocity between ions and electrons at altitudes between 90~km and 120~km. Here, the ions experience collisions detrimental to their drifts and their motion is subsequently restricted, while the electrons are able to drift freely owing to their much smaller gyroradius. The resulting polarization causes instability \cite{st.-mauriceNewNonlinearApproach2001}, when the drift speeds exceed $\sim 400$~m/s ($\sim 20$~mV/m).

\color{black}
Moreover, the Joule dissipation rate (heating) scales as the square of the electric field, so that the spatio-temporal variance of the field, and not its mean alone, enters the high-latitude energy budget. As a result, the quantity under investigation, the electric field, is a \color{black} vital quantity in space weather modeling \cite{codrescuImportanceEfieldVariability1995,rosenqvistMagnetosphericEnergyBudget2006,dengPossibleReasonsUnderestimating2007,knippDirectIndirectThermospheric2004}. \color{black} To complicate matters, the quantity is exceedingly hard to measure. \color{black} Orbiting satellites produce static one-dimensional slices through the plasma. \color{black} Ground-based coherent radars provide continuous coverage, but the standard convection products of the Super\textsc{darn} network are spherical-harmonic fits to line-of-sight velocities gathered across an entire hemisphere\cite{thomasStatisticalPatternsIonospheric2018} or inside a relatively large region \cite{rohelApplicationWideBeamTransmission2024,billettFastBorealisIonosphere2025}, procedures designed to recover constrained convection patterns, therefore suppressing variance below the scale of the fit. Conversely, the residuals to hemispheric fits retain substantial turbulent power \cite{cousinsStatisticalCharacteristicsSmallscale2012}. 

In the E region, as mentioned, coherent radar Doppler shifts are saturated by the FB instability and cannot recover the driving electric field directly \citep[see, e.g., ][]{fosterSimultaneousObservationsEregion2000,oppenheim_kinetic_2013}. However, recent advances in ionospheric \textit{radar tracking} measures the displacement of the scattering regions themselves, thereby bypassing the limitations imposed by saturation \cite{ivarsen_point-cloud_2024,ivarsen_deriving_2024}. The approach is naturally suited to extreme events: E-region irregularities require the field to exceed the instability threshold, facilitating the tracer precisely when the driving is strongest, which are the conditions when F-region coherent backscatter is least dependable \cite{currieSuperDARNBackscatterIntense2016,fioriExaminingPotentialSuper2018,blandSuperDARNRadarDerivedHF2018}, and when the \textit{E-region} carries most of the resulting dynamics \cite{themens_high_2024}. \color{black}

To explain how radar tracking can recover electric fields, we begin with a description of the auroral ionosphere, where \color{black} FB waves appear and disappear fast and in large quantities. \color{black} The sum total of these waves outline, in time and space, the structure of the driving electric field imposed by the magnetosphere \cite{ivarsen_characteristic_2025}. The short-lived, or \textit{ephemeral}, quality of the waves means that the occurrence of FB turbulence around aurorae act as tracers, tracing the shape and trajectory of moving electric field structures \cite{ivarsen_deriving_2024}. This was first exploited by \citeA{ivarsen_point-cloud_2024}, who provided electric field measurements around auroral arcs. This is enabled by the plasma being ``frozen into'' the magnetic field at altitudes higher than around 150~km, meaning that the motion of  electric field structures in the lower ionosphere tracks the convection electric field \cite{ivarsen_point-cloud_2024,ivarsen_deriving_2024}.

Building on the former, we present in this paper a comprehensive method to automatically identify and track clusters of radar echoes as they appear in the space above the Rabbit Lake observation station in Saskatchewan, Canada. This is achieved by amending the method presented by \citeA{ivarsen_point-cloud_2024} using modern tracking technologies commonly applied in aviation and defense contexts, as well as those applied to video-feed monitoring. Our paper concludes that the method can \color{black} routinely monitor electric field spikes and is capable of \color{black} detecting extreme ($>10,000$ m/s) events in the auroral ionosphere.

\section{Methods}

The suite we have implemented is a three-stage chain operating on \textsc{icebear} Level-2 echo data \cite{huyghebaertICEBEARAlldigitalBistatic2019,huyghebaertPropertiesICEBEARERegion2021}. Recently, \citeA{ivarsen_point-cloud_2024} implemented an algorithm to automatically track clutters, or clusters, of radar backscatter targets through time and space. The present paper improves on that procedure, with a functionality that can identify and follow the evolution of multiple, simultaneous targets.    We follow the evolving shapes from frame to frame in part by using a prediction based on previous time-frames.  We then proceed to  track their central position and finally infer a velocity \color{black} to the cluster. The tracker therefore maintains multiple simultaneous cluster identities, follows changes in their shapes, and records their creation and termination. \color{black}

The algorithm rests on four key steps, which are described in more detail below.  The four steps are,

\begin{enumerate}
\item Identification of  point-cloud footprints

\item Frame-to-frame association that allows for changes in the shape of clusters "within reason"


\item Handling the births, deaths, splits, and mergers of clusters, and  tracking their individual motions, with allowance for brief data gaps if necessary


 \item  Converting each tracked centroid trajectory into segmented velocities


\end{enumerate}

\subsection{Identification of point clouds footprints}

\color{black}
Each 3-s frame of \textsc{icebear} level-2 echo positions, advanced at 1-s cadence, is projected onto a local Cartesian plane and partitioned by the unsupervised learning algorithm \textsc{dbscan} \cite{esterDensitybasedAlgorithmDiscovering1996}, a density-based clustering algorithm, with minimum neighbourhood population of 15 and a neighbourhood radius $\epsilon$ determined per frame from the knee of the sorted $k^\text{th}$ nearest-neighbour distance distribution, $k = 15$, smoothed in time and clamped to [3, 200] km \cite{ivarsen_point-cloud_2024}. Unassigned points are discarded.

The boundary, or envelope, of each retained cluster is described by an `$\alpha$-shape' \citep{edelsbrunnerThreedimensionalAlphaShapes1994}, with $\alpha$ adaptively determined per cluster to the smallest value for which the shape forms a single region, giving the tightest non-convex outline that leaves the cluster connected.  \color{black}

\begin{figure}
    \centering
    \includegraphics[width=1.1\linewidth]{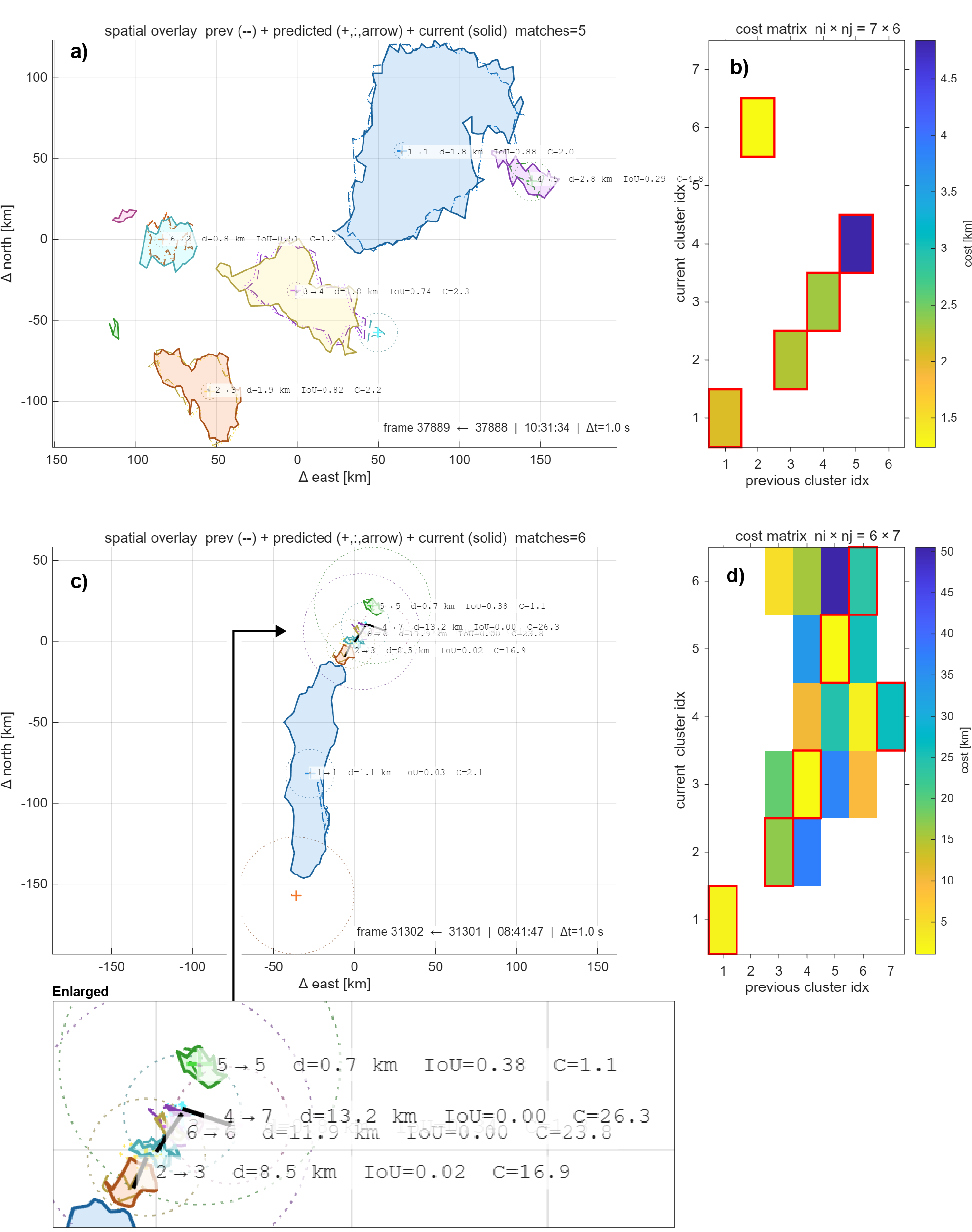}
    \caption{$\alpha$-shape association in two tracking situations (rows). Clusters are shown with their current $\alpha$-shape (solid outline, shaded area) compared to the previous, with gating radii indicated (36~km$\to$15~km). The cost matrix (panels b, d) allows targets to be tracked even in highly cluttered regions [the cluster of clusters around position $(0,0)$]; region shown in inset. \color{black} The data was collected at 10:31 UT (panels a, b) and 08:41 UT (panels c, d) on 17 July 2023. \color{black}}
    \label{fig:shapes}
\end{figure}

An example of $\alpha$-shapes used as a starting point is provided with color shaded areas in Figure \ref{fig:shapes}a). Recall that the clusters have areas of several km$^2$ (even exceeding 100 km$^2$ at times), that are filled, to various degree, with 3-m size Farley-Buneman (FB) unstable waves; \color{black} these waves scatter the radar signal, producing an observable echo. Since the FB instability strictly requires electron drifts to exceed the ion acoustic speed $C_s$, an echo detection signifies that the electric field has surpassed the threshold value of around 20~mV/m at that location. Clusters of such echoes, or threshold exceedance locations to be precise, constitute the tracked objects. \color{black}

For what comes next, we have to characterize  the progression in the $\alpha$-shapes as they are detected from one time-frame to the next, For this, we  use  an overlap tracker through a ratio defined as ``Intersection over Union" (IoU).  The numerator (intersection) is the area where the shapes from two successive time-frames overlap.  The denominator is the total covered area covered by both shapes put together.  That is to say, it is the sum of the areas of the two successive $\alpha$-shapes minus the area where they overlap.  

\subsection{Frame-to-frame association}

After individual clusters, or collections of clusters, have been identified, the algorithm looks at where each cluster has just been, and it then uses linear extrapolation from previous changes in position to predict where each cluster should appear next.  Because the $\alpha$-shapes undergo distortions from one time frame to the next, we cannot just track its centroid (center-of-mass). \color{black} The gate (search radius) is first considered; if the prospective cluster is outside of the search radius, association is impossible. Then, IoU is considered, using \textsc{matlab}'s \texttt{matchpairs} algorithm (the linear assignment problem). 
At this point,  \color{black} the algorithm (by way of an `associator') has  to handle $N_i$ clusters at frame $t$ and $N_j$ clusters at the most recent non-empty frame within a nine-frame look-back window (i.e., the tracker tolerates brief detection gaps). The associator uses labels to identify  each of the identified clusters.  The system ultimately reports the evolution of each of the labeled clusters whose evolution is measured.


Before a decision is made as to whether or not a cluster has been tracked, a cost-analysis has to be performed.  That cost-analysis  is based on a  linear assignment problem with an $N_i \times N_j$ cost matrix. As stated above, for each previous-frame cluster $j$ a prediction is made about its position at the next time frame $t$ by linear extrapolation using that cluster's instantaneous velocity. \color{black} The \textit{gating radius} is the maximum allowed distance between an observed centroid and its prediction. A newly born cluster has no velocity estimate, so its first association is gated on the raw centroid displacement, with $D_\text{gate} = [12,36]$~km. Once a velocity estimate exists, the gate is applied to the \textit{prediction residual} and tightened to $D_\text{predict} = [5,15]$~km, so that it constrains the frame-to-frame \textit{change} in velocity, rather than its magnitude. Clusters arising as splits inherit the parent velocity and are gated on the prediction residual from their first frame. The bracketed numbers delineate \textit{tight} and \textit{loose} schemes respectively. \color{black}

For all remaining pairs $(i, j)$ that are \textit{inside} the gating radius, we next  compute  the current cluster's \texttt{polyshape} (or $\alpha$-shape) against
the previous cluster's \texttt{polyshape} after its translation by the predicted
displacement \citep[see, e.g., ][ for optical implementations in industry applications]{bewleySimpleOnlineRealtime2016,wojkeSimpleOnlineRealtime2017}. 
Writing $d =
D_{\text{mat}}(i, j)$ for the centroid-to-prediction distance (in km), the
cost for the pair is defined by the equation,
\begin{equation}
    \text{Cost}(i, j) = d \cdot \bigl(1 + \beta \cdot (1 - \text{IoU})\bigr),
    \label{eq:cost}
\end{equation}
where the dimensionless weight $\beta$ controls the relative contribution of shape mismatch versus centroid distance. We set $\beta = 1$, the maximum-entropy choice in the absence of a calibration dataset, and a balance adopted implicitly by the SORT algorithm \citep[Simple Online and Realtime Tracking;][]{bewleySimpleOnlineRealtime2016}. With this choice, perfect shape continuity ($\text{IoU}=1$) reduces the cost to the kinematic distance $d$, while zero overlap doubles it to $2d$, which ensures that echo clusters that \textit{happen to be spatially close} but are in fact different objects will not disrupt the predicted evolution of any cluster. The construction is readily implemented as a live solver on a continuous data stream;   here, we apply it to historical data using \textsc{matlab}'s \texttt{matchpairs(Cost, D\_gate)}, which solves the assignment with the so-called Hungarian/Munkres algorithm \citep{kuhnHungarianMethodAssignment1955}.
In simple terms, this algorithm considers the $N_j$  mix of scorecards for observed clusters, and it finds the single global combination that gives the absolute best (i.e, smallest cost)  matching score for the entire set of observed clusters at time $t$.

Figure~\ref{fig:shapes}a, b) shows an example  of snapshots made by the associator, illustrating the process for a tidy, easy tracking situation, where each observed track is naturally assigned a predecessor. For a ``messy'', cluttered tracking situation, when many small clusters appear inside one another's gating radius, we refer to Figure~\ref{fig:shapes}c, d). 

\begin{figure}
    \centering
    \includegraphics[width=1.1\linewidth]{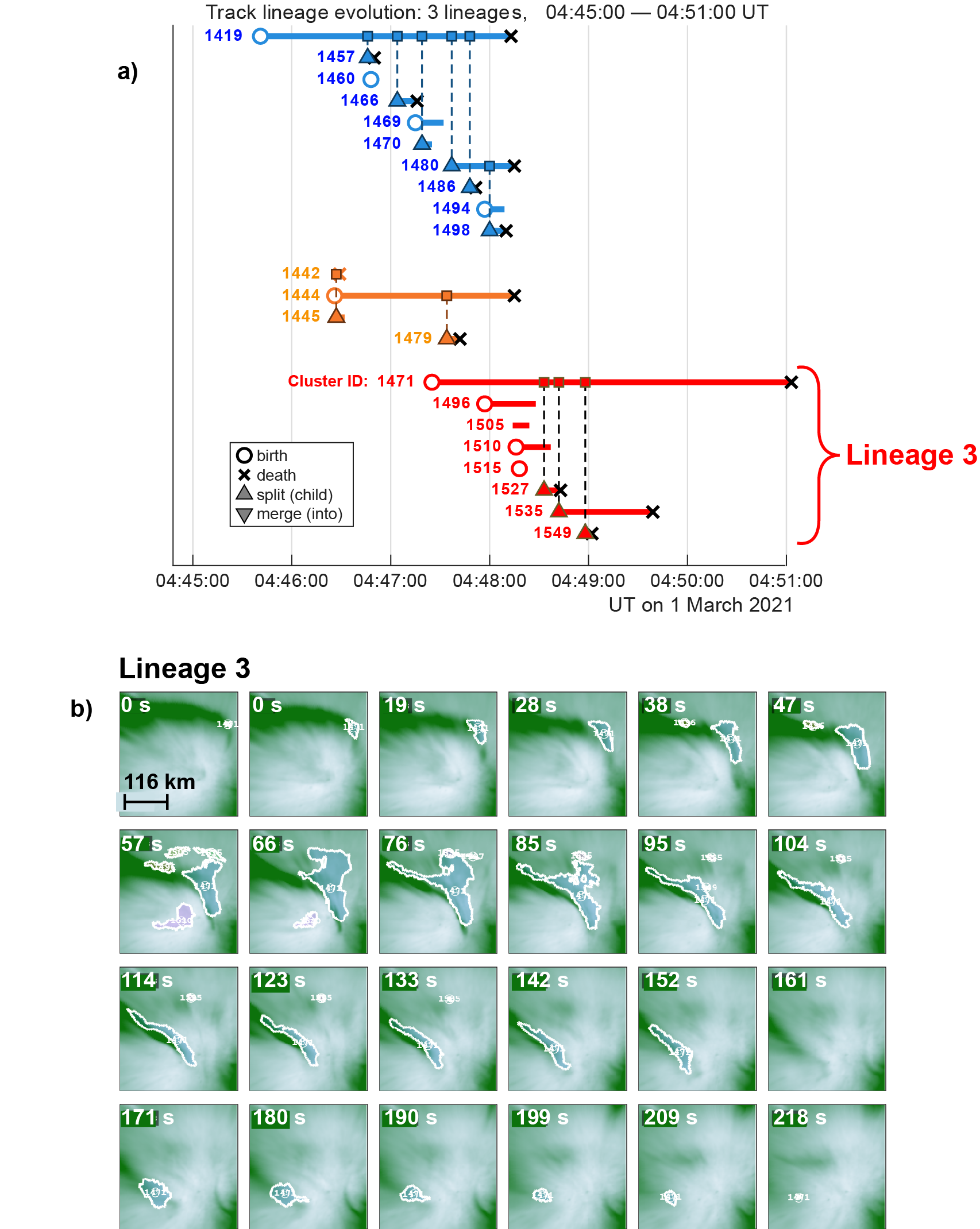}
    \caption{The three distinct lineages, or series of co-evolving clusters of radar echoes, recorded between 04:45 and 04:51~UT on 1 March 2021. Panel a) shows the lineages themselves; panel b) details `Lineage 3' in 24 snapshots. Clusters are plotted as $\alpha$-shapes overlaid on auroral images from the \textsc{tre}x \textsc{rgb} system \cite{gilliesApparentMotionSTEVE2020}. }
    \label{fig:lineage}
\end{figure}

\subsection{Births, deaths, mergers, and splits}

Even after the  foregoing assignment problem is taken care of,  there remain additional ways a cluster can fail to be clearly identified as a 1-to-1 match, meaning that the tracking ceases.
First of all, when a new shape appears (\textit{births}), if the intersection between it and a  shape from the previous time step covers more
than 5\% of the area of a cluster identified in the previous time step, the new  cluster is interpreted as
a \textit{split} product of the matched parent, giving the next, fresh, target an existing velocity estimation and corresponding narrow gating. This allows velocities to be monitored at different points over time and space, providing a flow field which is  consistent with the long tail of field strength events appearing and/or  quickly vanishing stochastic ensembles. This bookkeeping, \color{black} which allows \textit{mergers} and subsequent lineages to be tracked in time and space, \color{black} is borrowed from the field of extended-target tracking \citep[][and references therein]{granstromExtendedObjectTracking2017}. 

\subsection{Converting each tracked centroid trajectory into segmented velocities}
After the  full evolution from birth to death of each cluster has been documented, the \textit{trajectory} of the cluster-centroid (median location, or echo centre-of-mass) is treated as two scalar time series in geomagnetic displacement, one to the east and the other to the north, each in km, relative to the track's first point.  Each component is  modeled in piecewise fashion, and linearly in time. The algorithm splits the trajectory into intervals where a constant-velocity model \textit{fits}: a sliding window is extended forward one sample at a time until the linear correlation with time drops below a fixed threshold \color{black} ($\rho=2/3$). If no linear correlation exceeds $2/3$, or if the inferred speed is lower than $300$~m/s, \color{black} the tracking ceases over the region \citep[akin to online piecewise-linear approximations;][]{keoghOnlineAlgorithmSegmenting2001}. 

The method's central output, the speed of clusters in aggregate, is $|\mathbf{v}| = \sqrt{v_e^2 + v_n^2}$, where $e,n$ subscripts refer to eastward and northward velocities; this velocity is readily converted to electric fields if the local magnetic field vector is known (or estimated using a global model). Speed error is propagated by
standard quadrature:
\begin{equation}
    \sigma_{|\mathbf{v}|} =
    \frac{\sqrt{(v_e \sigma_e)^2 + (v_n \sigma_n)^2}}{|\mathbf{v}|}.
\end{equation}
In addition, the bearing, (or azimuth)  for the cluster's motion is retrieved from  $\theta = \mathrm{atan2}(v_e, v_n)$ wrapped to $[0^\circ,
360^\circ)$.

Figure~\ref{fig:lineage}a) offers a practical illustration of the overall process. The example is taken from between 04:45~UT and 04:51~UT on 1 March, 2021, when ``super-clusters'', or `lineages', appeared.   Figure~\ref{fig:lineage}b) shows the detailed evolution of a particularly interesting subset where a series of structures were closely following the poleward edge of an auroral arc for more than three minutes. During this interval, two clusters merged ($\sim47$~s mark), followed by a mass-merger ($\sim57$~s), a split-up ($\sim85$~s), and eventual decay.

\color{black}

\subsection{Difference from \citeA{ivarsen_point-cloud_2024}}

The scheme of \citeA{ivarsen_point-cloud_2024} used a nearest-neighbour approach, relying on rectangular ``bounding boxes'' for comparisons between frames. The present method differs in five respects. \textit{(1)} It handles an arbitrarily high number of clusters in each frame at once, so that association becomes an assignment problem solved globally rather than a sequence of independent nearest-neighbour choices. \textit{(2)} Candidates are selected based on the distance between a cluster's centroid and a position predicted from its own previous displacement, rather than on proximity alone, so that a fast-moving cluster is not accidentally assigned to a slower neighbour that happens to lie closer. \textit{(3)} The assignment cost combines that distance with the degree of $\alpha$-shape overlap between consecutive frames, which requires each cluster to carry an explicit and adaptive envelope (an $\alpha$-shape) rather than a rectangular bounding box. \textit{(4)} Clusters that split or merge are carried through the event with parent and child labels retained, so that a one-to-many transition no longer terminates a track; the resulting lineages can be characterized. Finally, \textit{(5)} each trajectory is divided into intervals over which a constant velocity describes the motion, yielding per-segment velocities in place of a single velocity per feature, so that changes in speed and direction \textit{within} a track are resolved rather than averaged away.

\color{black}

\section{Data}

\subsection{Case studies}

\begin{figure}
    \centering
    \includegraphics[width=\textwidth]{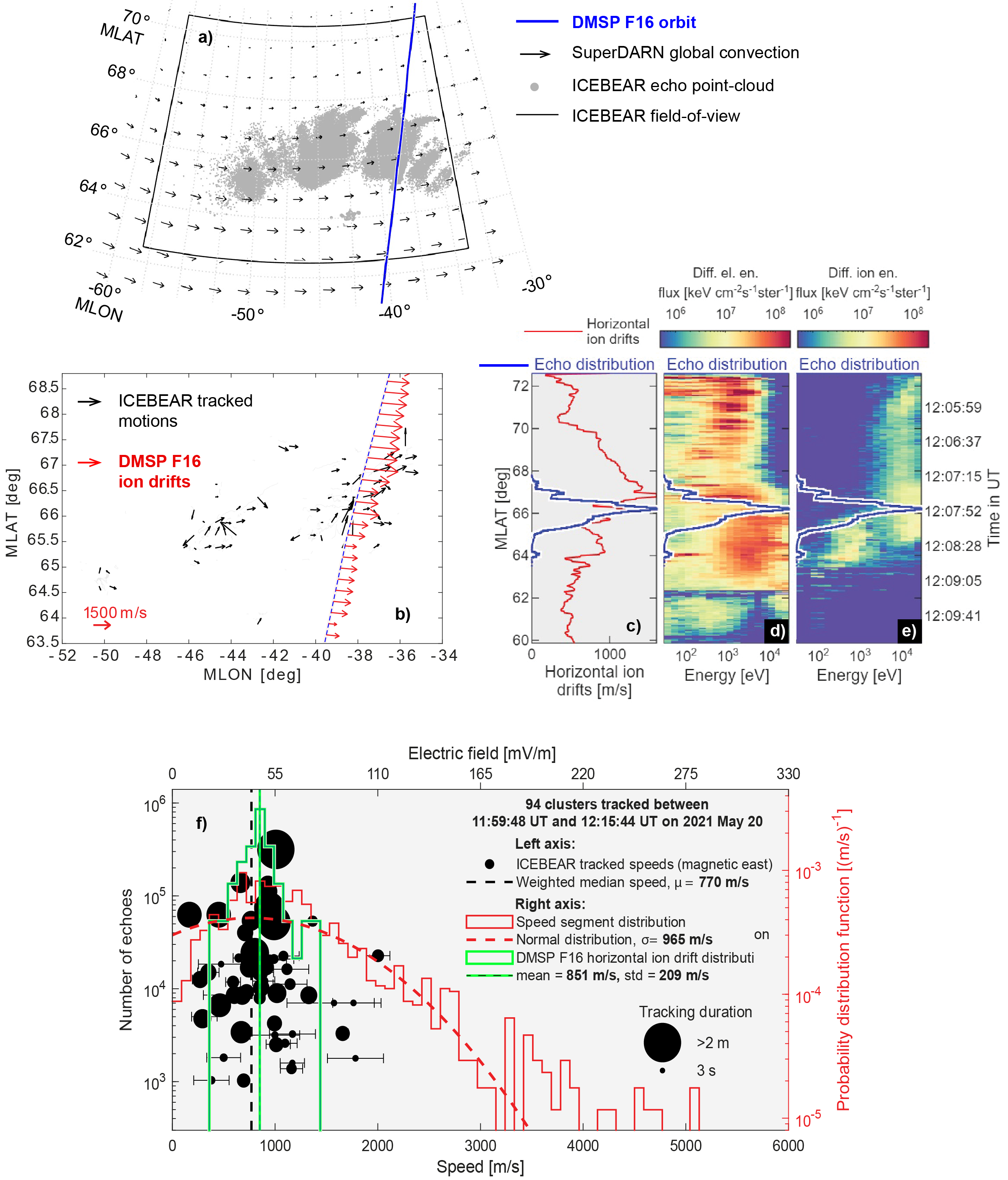}
    \caption{\textbf{Conjunction between DMSP F16 and \textsc{icebear} on 20 May 2021.} \textbf{Panel a)} shows DMSP F16 orbit (blue), superposed on a vector-field of Super\textsc{darn} global convection \cite{thomasStatisticalPatternsIonospheric2018} (black arrows) and \textsc{icebear} echo locations (gray dots). \textbf{Panel b)} shows the \textsc{icebear} tracked motions (black arrows) and the DMSP F16 horizontal ion drifts (red arrows). \textbf{Panel c)} shows the DMSP F16 horizontal ion drifts (red line), while \textbf{panels d) and e)} show the precipitating electron and ion differential energy flux, respectively, represented with a colorscale. In panels c--e), the latitudinal distribution of \textsc{icebear} echoes is superposed with a blue line. \textbf{Panel f)} shows, on the left $y$-axis, the distribution of tracked \textsc{icebear} clusters (black circles), taking the eastward component only, with a dashed black line giving the weighted median speed (770~m/s). On the right $y$-axis, the distributions in \textsc{icebear} per-frame displacements is shown in a solid red line, with a Gaussian fit displayed with a dashed red line ($\sigma=965$~m/s). The distribution in DMSP F16 ion drifts is shown with a green line, with a mean of 851~m/s and a standard deviation of 209~m/s.   }
    \label{fig:dmsp}
\end{figure}

\begin{figure}
    \centering
    \includegraphics[width=0.9\linewidth]{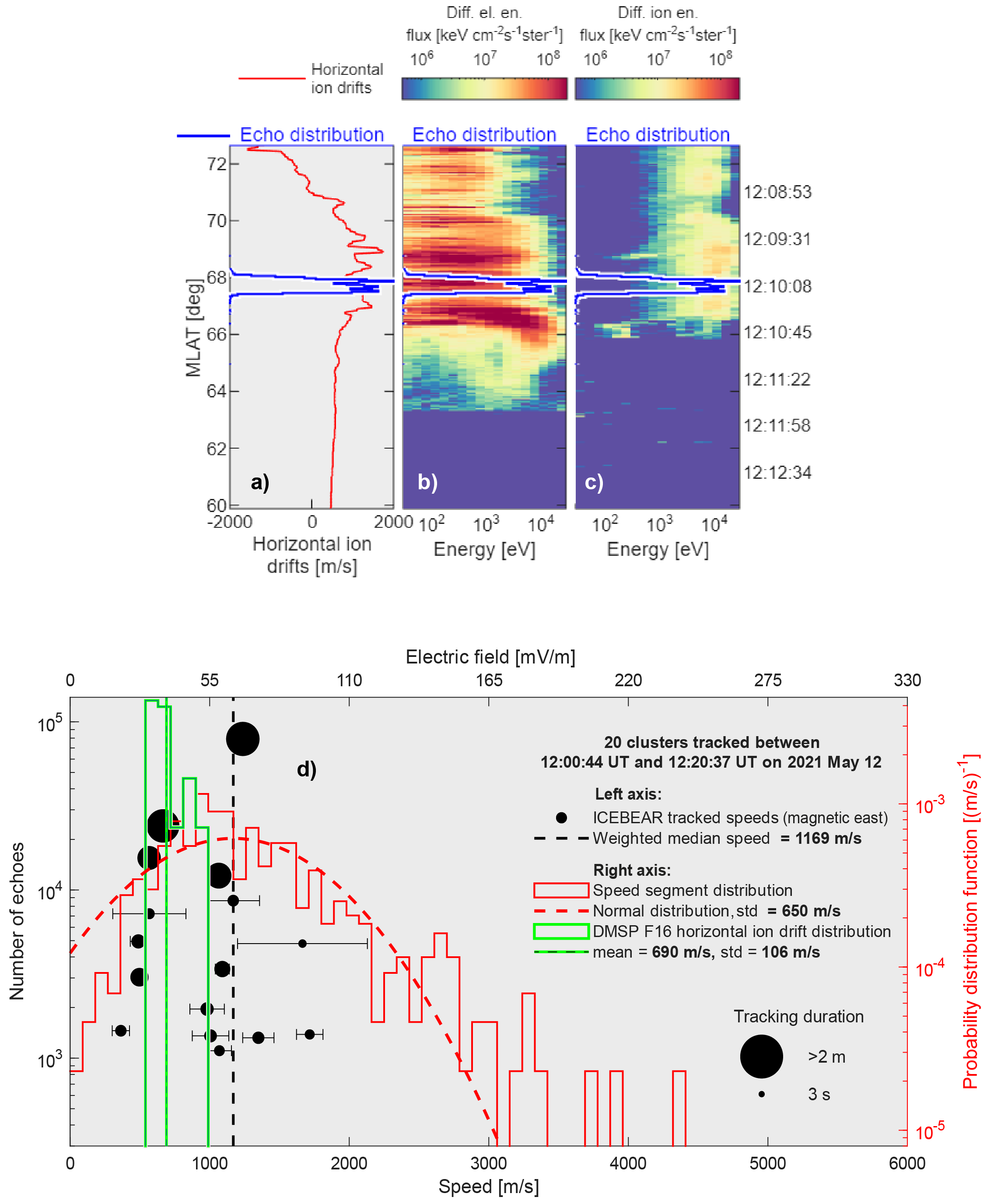}
    \caption{A similar conjunction to Figure~\ref{fig:dmsp}, \color{black} occuring around 12:10 UT on 12 May 2021, \color{black} with panels a--c) showing the ion drift and particle precipitation data from DMSP~F16 (echo distributions overlaid in blue), and panel d) showing the window variability in tracked speeds (black circles), individual per-frame displacement speeds (red line), and DMSP~F16 ion drifts (green line).
    }
    \label{fig:dmsp2}
\end{figure}

The methods of Section~2 turn each radar-aurora cluster into a labeled \textit{trajectory}, or track, and, via the magnetic-field projection of Section~2.4, they turn into a local electric-field estimate \citep{ivarsen_deriving_2024}. This section brings four space-ground conjunctions to bear on the above interpretation. Figures~\ref{fig:dmsp} (20 May 2021) and \ref{fig:dmsp2} (12 May 2021)  anchor the tracked motions to in-situ ion drifts measured by DMSP~F16, confirming the proxy relation under standard auroral driving. Figures~\ref{fig:metop}--\ref{fig:noaa} (10 May 2024) apply the same proxy during the early main phase of the recent G5 super-storm, with observations made on closed field-lines equatorward of the cusp \citep[the conjunctions were first reported in][]{ivarsen_eastward_2025}. During the two super-storm conjunctions, we detect 10 clusters moving faster than 6000~m/s, of which the fastest was a 5-second event with a fitted speed of $11{,}240\pm660$~m/s.

During the 16-minute interval between 11:59:48 and 12:15:44~UT on 20 May 2021, \textsc{icebear} resolved 94 clusters of sufficient duration to enter the segmented-velocity analysis (Figure~\ref{fig:dmsp}f). The weighted-median segmented speed was $770$~m/s; the speed-segment distribution is well described by a Gaussian of $\sigma = 965$~m/s, save for the long tail. The DMSP F16 horizontal ion-drift moments collected over the same interval averaged $851$~m/s with a $1\sigma$ scatter of $209$~m/s. The two distributions agree on the mode, but the \textsc{icebear} distribution is considerably broader, owing to a large number of in-cluster displacements (whose implied speed distribution is represented by the solid red line in Figure~\ref{fig:dmsp}f). These are averaged out in the quantified tracks --- witness that only five tracks are faster than 1400~m/s and the rest of the tracks exhibit speeds that are wholly enclosed by green histogram (the F-region ion drifts). The long tail of segment speeds in Figure~\ref{fig:dmsp}f) represent small-scale variability, captured by \textsc{icebear}'s exceedingly high spatio-temporal resolution.  

All the 94 tracked target motions are on display as a whole in Figure~\ref{fig:dmsp}b), showing arrows as a result from spatio-temporal averages of the tracking segments. Whereas the satellite's one-dimensional slice gives the impression of a steady eastward `flow', \textsc{icebear} can observe `counter-streaming' behaviour in this \textit{flow}, which at times overwhelm the ambient field in any which direction, but we note a slight preference for the eastward direction. In Figure~\ref{fig:dmsp}c--e), we note that the radar echoes were observed on the poleward side of a diffuse, high-energy auroral patch, coinciding with the precipice of low-energy precipitation, as well as the latitude of the peak ambient field-strength.

Looking at the average and the distributions in Figure~\ref{fig:dmsp}f), we observe that the agreement between the green distribution and the black circles  extends the validation of \citet{ivarsen_point-cloud_2024,ivarsen_deriving_2024,ivarsen_eastward_2025-1}, who favourably compared tracked radar motions to ion drifts inferred from coincident Swarm, DMSP, and Super\textsc{darn} observations. A similar conjunction between \textsc{icebear} and DMSP~F16 that occurred some eight days earlier, on 12 May 2021, is consistent with the emerging picture (Figure~\ref{fig:dmsp2}). 

\subsection{Statistical validation}

\color{black}
The two conjunctions of Section 3.1 sample a few hundred tracked clusters in the ordinary auroral ionosphere; we now place them within the full distribution of tracked speeds observed during four years of radar operation, compared against the European Space Agency's Swarm~A cross-track ion drifts in the same region of space, over the same time interval. \color{black}

\begin{figure}
    \centering
    \includegraphics[width=0.79\linewidth]{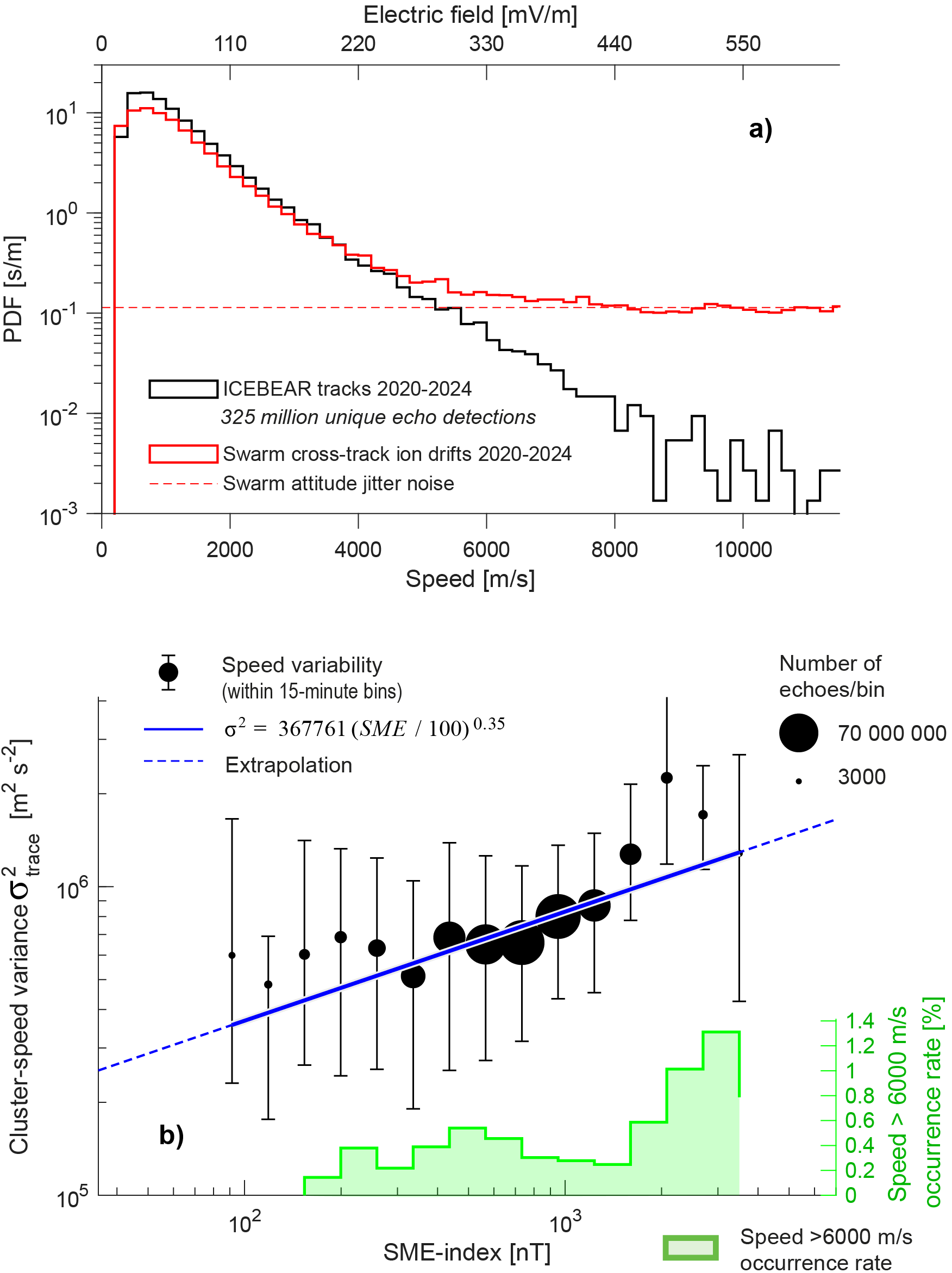}
    \caption{\textbf{Panel a)} shows, in a solid black line, the distribution of 74,517 clusters tracked by \textsc{icebear} during geomagnetically disturbed conditions (SME-index $>150$~nT), \color{black} Using a \textit{tight} tracking scheme ($D_\text{gate} = 12$~km and $D_\text{predict} = 5$~km). \color{black} The distribution of ion drifts speed derived from Swarm~A EFI horizontal moments, during the same conditions and inside the auroral region (18h$<$MLT$<$6h, 61$^\circ<$MLAT$<69^\circ$), is shown with a solid red line. A dashed red line indicates the noise limitation caused by spacecraft attitude jitter and anomalous spacecraft potential and calibration drift \cite{knudsenThermalIonImagers2017}. \textbf{Panel b)} shows the same dataset, this time binning all the tracked clusters into 4190 bins of 15-minute length (for a total of 1047.5 hours of radar-time), \textit{then} binning (for the second time) the \textit{standard deviations}, $\sigma$, of the cluster speeds inside each window, for the 24 geomagnetic activity bins. Black circles with errorbars denote average $\sigma$ in each geomagnetic activity bin, and a solid blue line shows the result of a log-log linear regression (Pearson correlation $=0.84$). In green, on the right $y$-axis, the occurrence rate for speeds $>6000$~m/s is shown for each geomagnetic activity bin.}
    \label{fig:stats} 
\end{figure}

The four-year \textsc{icebear} cluster-speed distribution is shown in Figure~\ref{fig:stats} against the corresponding distribution of Swarm~A's electric field instrument \cite[EFI; ][]{knudsenThermalIonImagers2017} cross-track ion-drift moments over the same period. Swarm~A is a polar-orbiting spacecraft at an altitude of around 450~km \cite{friis-christensenSwarmConstellationStudy2006}. Between $\sim 300$ and $\sim 4000$~m/s the two distributions match within a factor of two in PDF, a strong indication that the radar-aurora tracking is sampling the \textit{same} physical drift population that the \textit{in-situ} cross-track ion-drift measurement samples \citep[which was established for an earlier iteration of the algorithm by ][]{ivarsen_point-cloud_2024,ivarsen_deriving_2024}. Above $\sim 5000$~m/s the Swarm distribution flattens at $\text{PDF} \approx 10^{-1}$~s/m, in the regime where the cross-track ion-drift moment is dominated by attitude-jitter and calibration-drift contributions \citep{knudsenThermalIonImagers2017}. The \textsc{icebear} distribution continues with an apparent powerlaw through this threshold and beyond, reaching the bin-count noise floor at $\sim 9000$~m/s. Cluster~3624 (Figure~\ref{fig:metop}f, see also Figures~\ref{fig:metop2} and \ref{fig:noaa}f) belong to the upper-tail sample of this distribution.

Figure~\ref{fig:stats}b) shows the geomagnetic activity trends in the track dataset, via the speed variability $\sigma$ (within running 15-minute bins), compared to the occurrence rate of superfast ($>6000$~m/s) clusters within each geomagnetic activity bin, where we use the SuperMAG auroral electrojet index \citep[SME; ][]{newellEvaluationSuperMAGAuroral2011} to quantify geomagnetic activity. A solid blue line shows the result of fitting a log-log linear fit (a power law) to the black data, which we shall discuss in Section~4.2. 


\subsection{super-storm sample}

\color{black}

We now turn to two conjunctions recorded during the main phase of the G5 super-storm of 10 May 2024, on the dayside near the equatorward edge of the cusp \citep[see ][]{ivarsen_eastward_2025,ivarsen_eastward_2025-1,madhanakumar_co-ordinated_2025}, which populate the upper tail of the distribution presented in Section 3.2.
\color{black}

\begin{figure}
    \centering
    \includegraphics[width=0.95\linewidth]{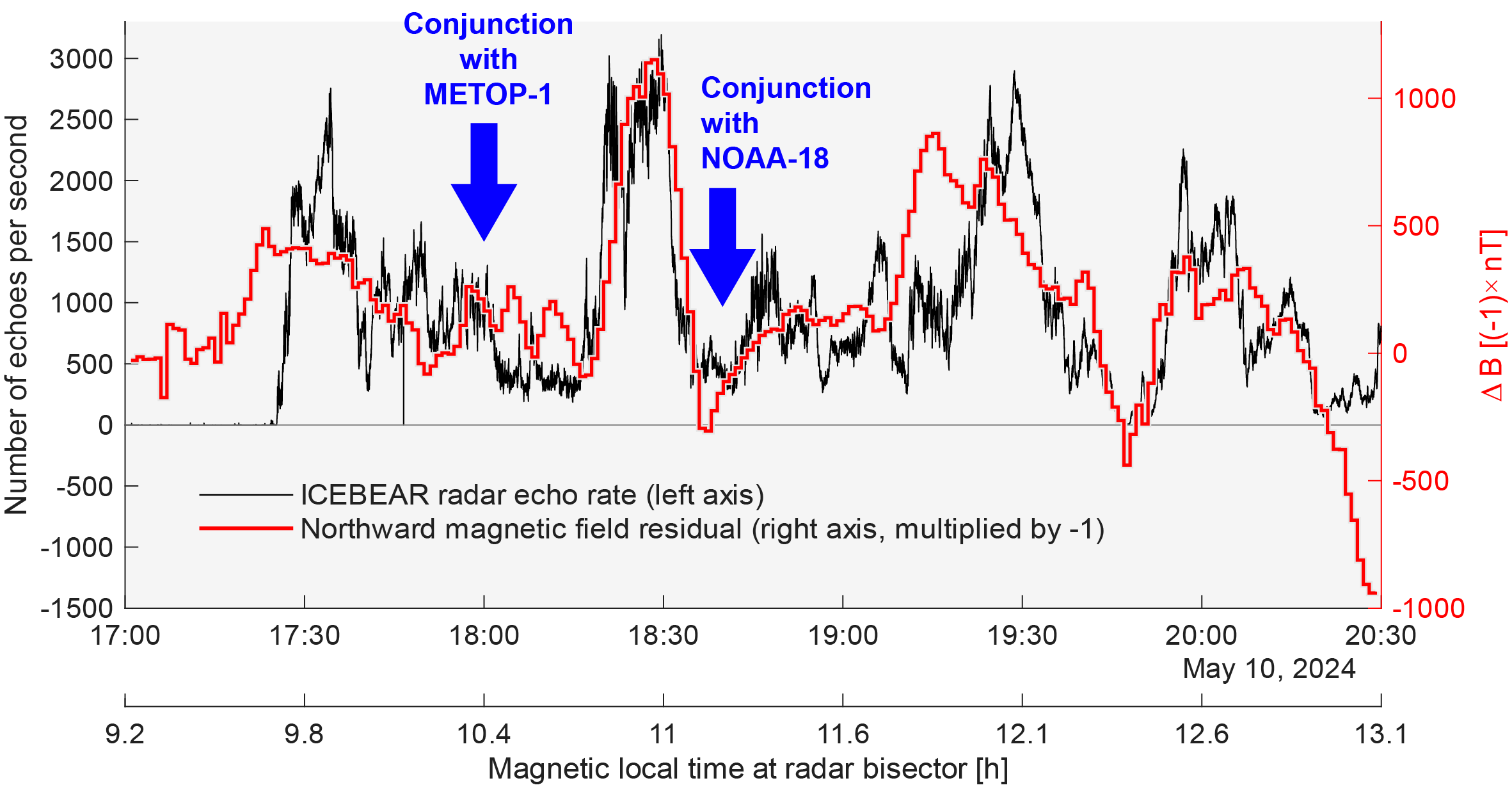}
    \caption{The rate of received \textsc{icebear} echoes (in number of echoes per second), shown in black line and on the left $y$-axis, compared with the northward magnetic field component observed by the Rabbit Lake  magnetotometer \cite{mannUpgradedCARISMAMagnetometer2008}, with the values inverted, and plotted on the right $y$-axis, with a red line. The timing of the two conjunctions between \textsc{icebear} and \textsc{metop}-1 and between \textsc{icebear} and \textsc{noaa}-18 are indicated, and the magnetic local time (MLT) of the radar bisector is indicated on the bottom $x$-axis.}
    \label{fig:indices}
\end{figure}

We analyze two space-ground conjunctions that took place around 18:00~UT and 18:40~UT on 10 May 2024. Figure~\ref{fig:indices} shows, in red line, \color{black} the northward magnetic perturbation recorded at Rabbit Lake with its sign inverted, which reflects the intensity of the overhead Hall currents, and which we take to represent the local development of this super-storm. \color{black} Superposed are the rate of echo detection, as well as the timing of our two fortuitous conjunctions.

\begin{figure}
    \centering
    \includegraphics[width=\textwidth]{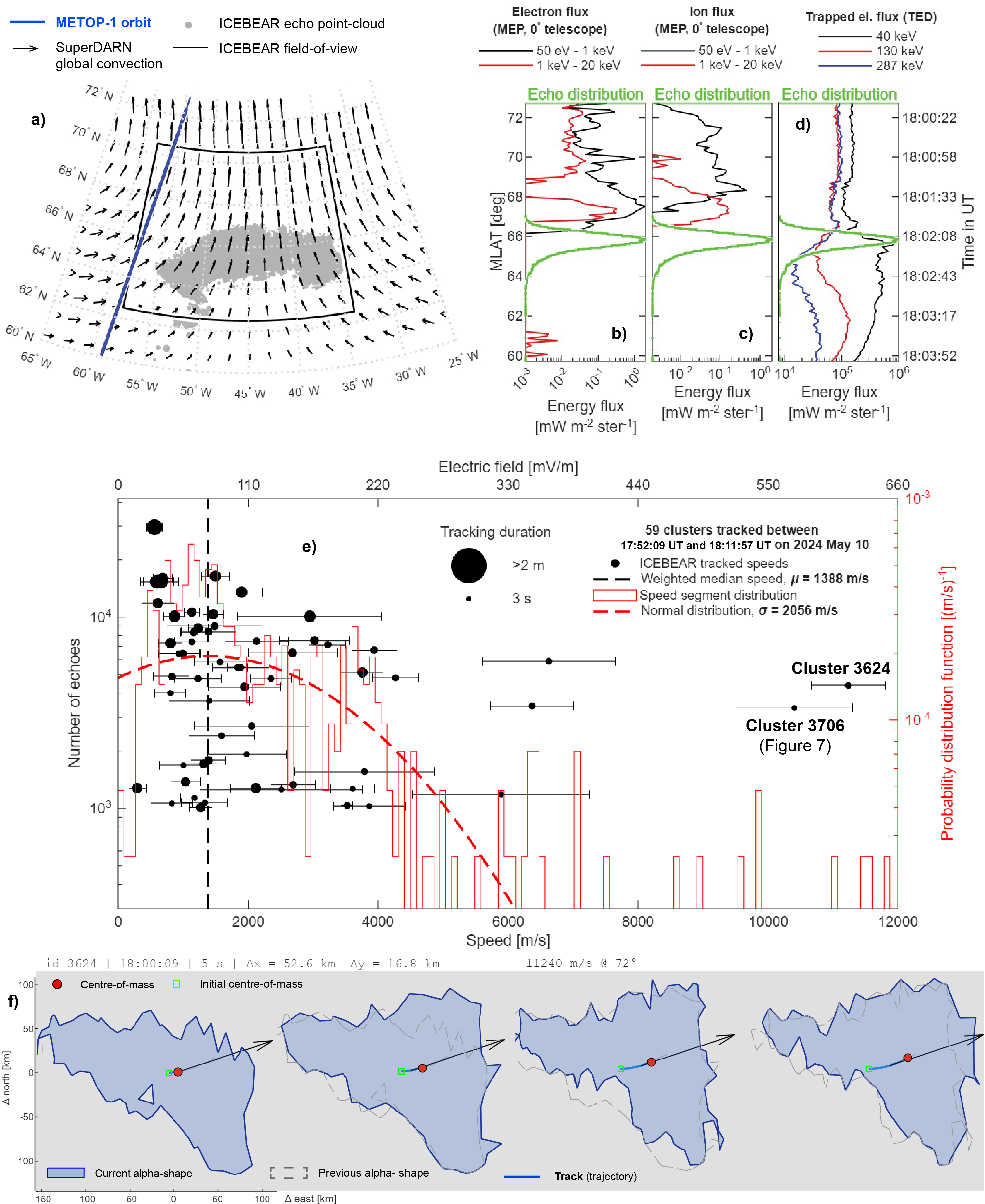}
    \caption{\textbf{Conjunction between METOP-1 and \textsc{icebear} at 18:00~UT on 10 May 2024.} \textbf{Panel a)} shows \textsc{metop}-1's orbit (blue), superposed on a vector-field of Super\textsc{darn} global convection \cite{thomasStatisticalPatternsIonospheric2018} (black arrows) and \textsc{icebear} echo locations (gray dots).  \textbf{Panels b) and c)} show the electron and ion fluxes, respectively, measured by \textsc{metop}-1 (medium-energy, $0^\circ$ telescopes), while \textbf{panel d)} shows the trapped electron fluxes.  In panels b--d), the latitudinal distribution of \textsc{icebear} echoes is superposed with a green line. \textbf{Panel e)} shows, on the left $y$-axis, the distribution of tracked \textsc{icebear} clusters (black circles) that were tracked during a 20-minute interval centred on the conjunction, with a dashed black line giving the weighted median speed (1388~m/s). On the right $y$-axis, the distributions in \textsc{icebear} velocity segments is shown in a solid red line, with a Gaussian fit displayed with a dashed red line ($\sigma=2056$~m/s). Cluster 3624, whose speed was determined to be $11{,}240\pm660$~m/s, is indicated, and its evolution is shown in detail in \textbf{panel f)}, showing the displacement of the identified $\alpha$-shape (see Section~2).
    }
    \label{fig:metop}
\end{figure}

\begin{figure}
    \centering
    \includegraphics[width=\textwidth]{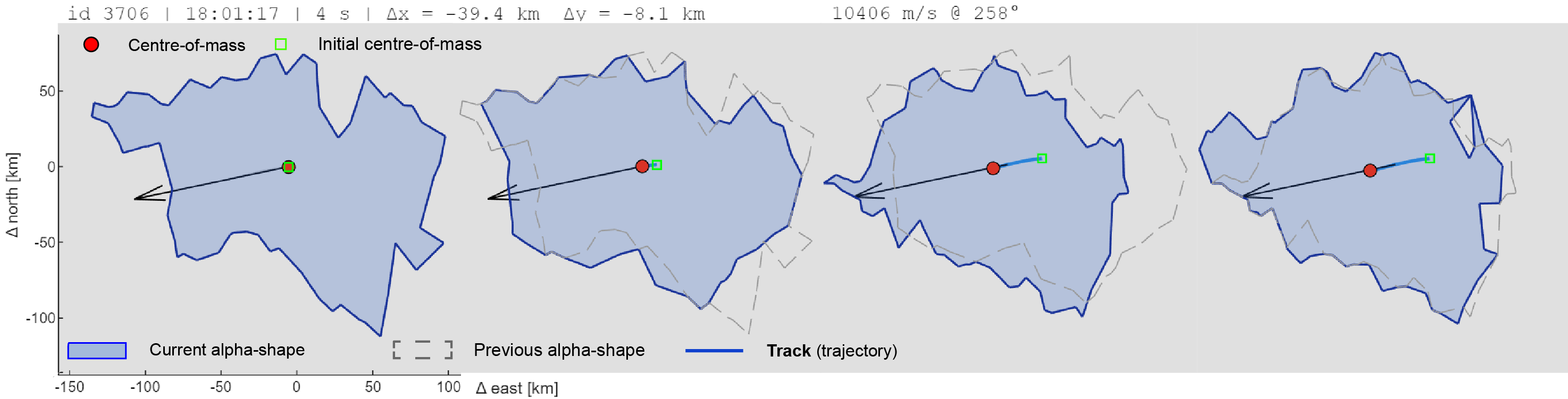}
    \caption{The evolution of Cluster 3706, whose speed was determined to be $10{,}406\pm900$~m/s, in four snapshots, showing the displacement of the identified $\alpha$-shape (see Section~2 and Figure~\ref{fig:metop}f).}
    \label{fig:metop2}
\end{figure}

Figure~\ref{fig:metop} details the \textsc{metop}-1 conjunction, which took place at $\sim$18:00~UT on 10 May 2024, during the main phase of the G5 super-storm of that date, at or near the equatorward edge of the dayside cusp (a signature of the cusp itself was detected by the satellite \textsc{noaa}-18 some forty minutes later; see Figure~\ref{fig:noaa} below). \textsc{metop}-1 is a climate satellite operated by \textsc{eumetsat}, with an instrumentation similar to that of the United States \textsc{noaa} satellites, capable of measuring precipitating particles through a number of telescopes.   \textsc{icebear} resolved 59 clusters in the 20-minute interval centred on the conjunction. The cluster-speed distribution (Figure~\ref{fig:metop}e) had a weighted median of $1388$~m/s and a segment-fit $\sigma = 2056$~m/s, both larger by a factor of approximately two, compared to the values measured during the 2021 conjunction in Figure~\ref{fig:dmsp}. The MEP precipitating-particle channels in Figure~\ref{fig:metop}b, c) show considerable energy fluxes in both the $50$~eV--$1$~keV electron and ion channels, co-located in \textsc{mlat} with the bulk of the \textsc{icebear} echo distribution; the TED channel in panel d) shows substantial trapped electron flux at $40$, $130$, and $287$~keV across the same \textsc{mlat} band. This combination places the observed radar aurora on \textit{closed field-lines} just inside the open-closed boundary of the dayside auroral oval, consistent with the configuration documented for this event by \cite{ivarsen_eastward_2025}, and with the similar conditions observed during the 23 April 2023 storm \cite{ivarsen_eastward_2025-1}. We note that Figure 2D in \citeA{themens_high_2024} corroborate the foregoing descriptions of the geospatial context of our observations.

\begin{figure}
    \centering
    \includegraphics[width=\textwidth]{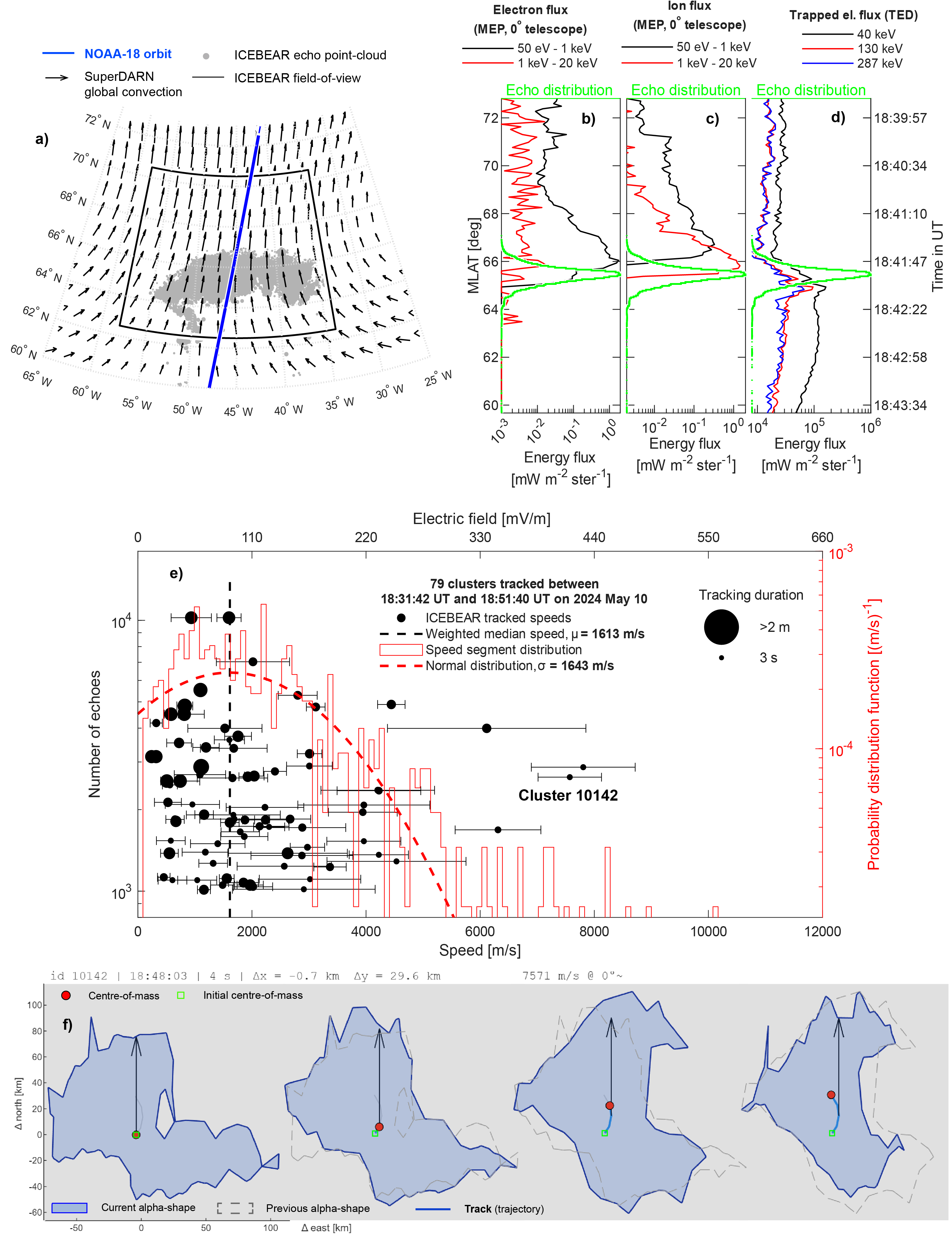}
    \caption{
    \textbf{Conjunction between NOAA-18 and \textsc{icebear} at around 18:40~UT on 10 May 2024.} \textbf{Panel a)} shows \textsc{noaa}-18's orbit (blue), superposed on a vector-field of Super\textsc{darn} global convection \cite{thomasStatisticalPatternsIonospheric2018} (black arrows) and \textsc{icebear} echo locations (gray dots).  \textbf{Panels b) and c)} show the electron and ion fluxes measured by \textsc{noaa}-18 (medium-energy, $0^\circ$ telescopes), while \textbf{panel d)} shows the trapped electron fluxes.  In panels b--d), the latitudinal distribution of \textsc{icebear} echoes is superposed with a green line. \textbf{Panel e)} shows the distribution of tracked \textsc{icebear} clusters (black circles) around the time of the conjunction. On the right $y$-axis, the distributions in \textsc{icebear} velocity segments is shown in a solid red line, with a Gaussian fit displayed with a dashed red line ($\sigma=1643$~m/s). Cluster 10142 is indicated, whose evolution is shown in detail in \textbf{panel f)}, akin to Figure~\ref{fig:metop}.
    }
    \label{fig:noaa}
\end{figure}

A particular cluster within the interval, labeled by the tracking algorithm as Cluster~3624 (Figure~\ref{fig:metop}f), had a fitted speed of $11{,}240\pm660$~m/s at an azimuth of $72^\circ$ (clockwise from magnetic north) over a tracked duration of 5 seconds. The centroid translated $\Delta x \approx 53$~km eastward and $\Delta y \approx 17$~km northward over that interval. Converting via $|\mathbf{E}| = |\mathbf{v}|\,|\mathbf{B}|$ with $|\mathbf{B}|$ from \textsc{igrf}-13 at the centroid position implies an electric-field magnitude of $\approx 560$~mV/m (recall that this value is obtained from the slope of the sliding-window piecewise-linear regression over the cluster's five samples). The $\alpha$-shape sequence in Figure~\ref{fig:metop}f) preserves the cluster's overall shape throughout the short evolution, with the centroid path straight, and the kinematic prediction at each step is landing inside the next observed $\alpha$-shape; the extreme trajectory is accurately measured using the tracking algorithm. \color{black} A second cluster in the same interval, Cluster 3706, was fitted at $10,406\pm900$~m/s and is shown in Figure~\ref{fig:metop2}. \color{black}

Some forty minutes later, the United States' climate satellite \textsc{noaa}-18 orbited through the same region, this time traversing the ionospheric \textit{cusp} \cite{ivarsen_eastward_2025}; Figure~\ref{fig:noaa}b--d) show a characteristic signature; displaying \textit{(1)} a strong low-energy electron flux, \textit{(2)} a weak or non-existent high-energy electron flux, \textit{(3)} a strong flux of ions, and \textit{(4)} a magnetic local time of 11h (see Figure~\ref{fig:indices}). Together, the observations \textit{(1-4)} mean that the satellite threaded flux tubes belonging to the ionospheric cusp \citep{newellCuspCleftBoundary1988,newellHemisphericalAsymmetryCusp1988,ivarsenGNSSScintillationsCusp2023}. The distribution of \textsc{icebear} echoes in Figure~\ref{fig:noaa}a--d) then demonstrate that the active echo region occurred on closed field-lines, on or near the equatorward edge of the cusp. While the tracked speed variability during this second super-storm conjunction is now slightly lower ($\sigma=1643~m/s$ vs $\sigma=2056$~m/s before), we observe four clusters moving faster than 6000~m/s, and we detail the evolution of one of those clusters in Figure~\ref{fig:noaa}f). 

Both the May 2024 conjunctions occurred on \textit{closed field lines} \cite[Figures~\ref{fig:metop}d and \ref{fig:noaa}d; see also][]{ivarsen_eastward_2025}, with widespread particle precipitation in the vicinity of the echo distribution \citep[see also ][]{themens_high_2024,madhanakumar_co-ordinated_2025}. This matches the configuration documented for dayside, eastward transients on closed field-lines \cite{ivarsen_eastward_2025,ivarsen_eastward_2025-1}, and we therefore attribute the measured field-variability to diffuse, dayside particle precipitation \cite{spasojevic_drivers_2010,ni_chorus_2014}, which is shown to be present in the region in Figure~\ref{fig:metop}b, c).

The foregoing exposition makes the $560$~mV/m value recorded in Figure~\ref{fig:metop}f) (as well as Figures~\ref{fig:metop2} and \ref{fig:noaa}f) remarkable. The reported speeds exceeds the documented \textsc{steve} distribution, which corresponds to field strengths of $100$--$200$~mV/m \citep{gilliesApparentMotionSTEVE2020,mishinInnerStructureSTEVELinked2023}, and it meets or exceeds the most extreme reported \textsc{said} events at $\sim 400$~mV/m \cite{andersonMultisatelliteObservationsRapid2001}.


\section{Discussion \& Summary}


We have tracked clusters of E-region radar aurora through time and space with a Hungarian-association scheme acting on $\alpha$-shape footprints, themselves enclosing clusters of radar echoes in space, and we have reduced each tracked cluster's trajectory to piecewise-linear \textit{velocities}. 

\color{black}

\subsection{The nature of the observations}

However, before interpreting these velocities, we must briefly expound what exactly the radar is tracking. A set of radar echoes are simply locations within the observable volume where the FB threshold is exceeded by the $\boldsymbol{E}\times\boldsymbol{B}$-drifts. A cluster, represented by its $\alpha$-shape, is therefore not a coherent object, but rather a geometric envelope drawn around a point-set of FB threshold-exceedances. Furthermore, the regulation of the FB instability is rapid: at 3~m, the linear growth rates are of order milliseconds and the diffusive decay time tens of milliseconds \citep{fejer_ionospheric_1980}, so the meter-scale population adjusts quasi-instantaneously, with the turbulent volumes decaying in 1-2~s \cite{prikryl_doppler_1988,prikryl_evidence_1990}. This rapid regulation is what enables us to track the centre-of-mass of the echo set.

Next, we must point out that the $\alpha$-shape envelope and its echo population are separate quantities, and a cluster may consequently be very large without being densely filled. To characterize this degree of filling, or ``packing'', which will facilitate a direct comparison of clusters with differing sizes and lifetimes, we normalize the cluster's echo count $N$ by the $\alpha$-shape area $A$ and the track duration $\tau$. The resulting quantity, $N/(A\tau)$, is the mean rate at which echoes are detected per unit area within a cluster, which, under the reading that an echo set are points of FB threshold-exceedances, becomes the first-order intensity of the point process \citep[see, e.g., ][]{daleyIntroductionTheoryPoint2006,coxStatisticalMethodsConnected1955,kostinskiSpatialDistributionCloud2000}.

\begin{figure}
    \centering
    \includegraphics[width=.95\textwidth]{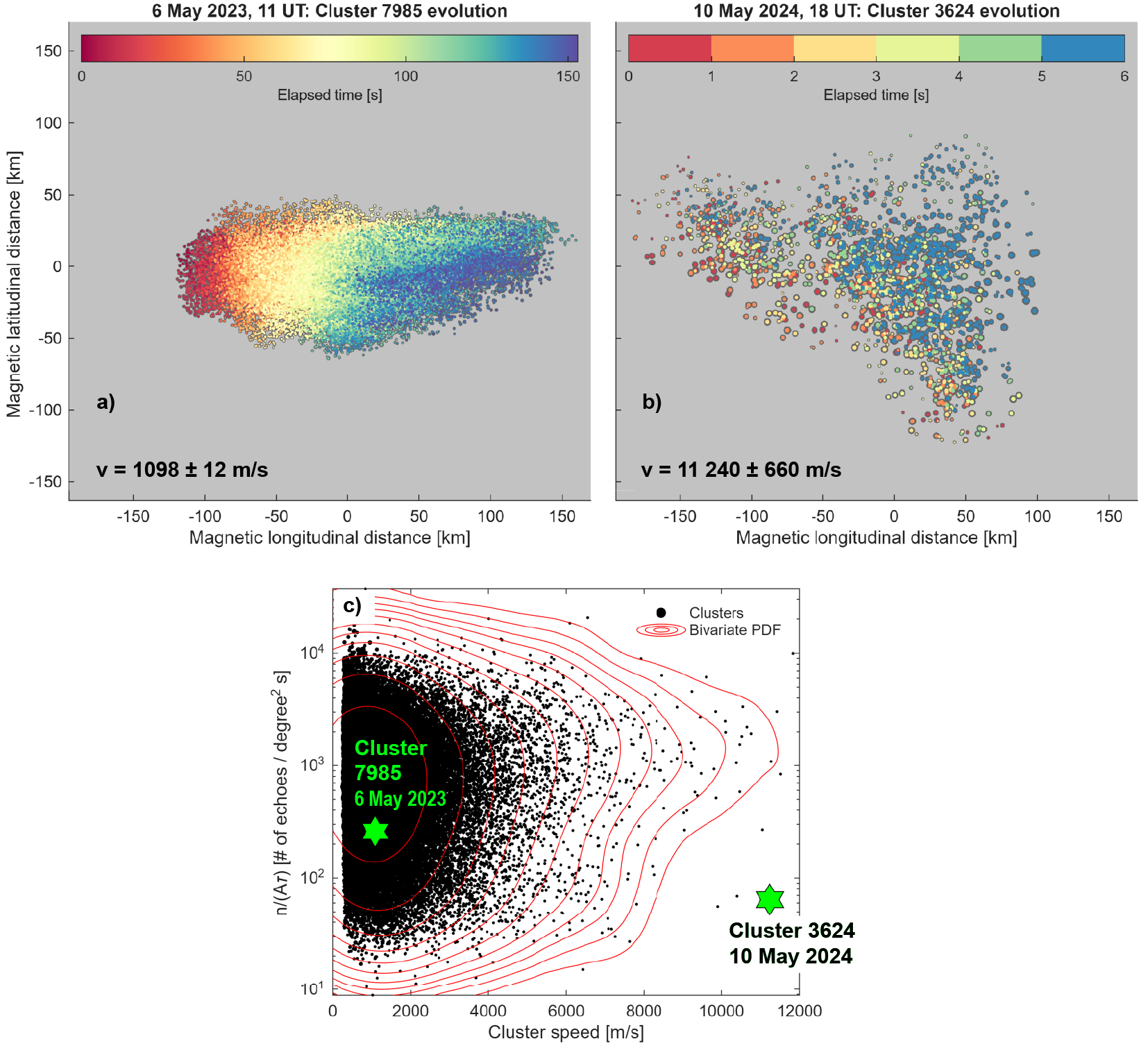}
    \caption{
    \color{black}
    The detailed evolution of the tracked echoes contained within cluster 7985 on 6 May 2023 \textbf{(panel a)} and cluster 3624 on 10 May 2024 (\textbf{panel b}, see also Figure~\ref{fig:metop}f). Individual echoes are shown as circle datapoints, color-coded by the time elapsed from the beginning of the track. \textbf{Panel c)} shows the distribution, in 74,517 tracked clusters, of echo counts normalized by $\alpha$-shape area (in degrees squared) and track duration (in seconds); clusters 7985 and 3624 from panels a) and b) are indicated. Linearly spaced contour lines of the estimated bivariate probability density function (PDF) are overlaid in thin, red lines. In panel c), four clusters whose velocities exceed 12,000~m/s are excluded (corresponding to probability density function values lower than $10^{-3}$~s/m; see Figure~\ref{fig:stats}a).
    \color{black}
    }
    \label{fig:rev}
\end{figure}

Figure~\ref{fig:rev} shows two examples, both of which move in the magnetic east direction, and one of which is the large, extremely fast ($>10,000$~m/s) cluster ``3624'', tracked on 10 May 2024 (Figure~\ref{fig:metop}f).  Figure~\ref{fig:rev}a, b) show individual echo locations color-coded by tracking-time; Figure~\ref{fig:rev}c) shows $N/(A\tau)$ for the entire tracked database. We observe that the large $\alpha$-shape of cluster 3624 on 10 May 2024 contains only some 750 echoes per frame, making the $\alpha$-shape an envelope drawn around a \textit{sparse set}: the field within the region is highly structured and the echo set marks where the field exceeds the $\sim20$~mV/m FB threshold. We surmise that the mean-field inside this large region is in fact at or below threshold, with a variance dominated by the extreme spikes of the structured field within. To see why this assertion is correct, consider the alternative: A field of $560$~mV/m amplitude sustained uniformly across cluster 3624's large envelope would require an electric potential difference across the area comparable to the entire storm-time cross-polar-cap potential. The field within the envelope must therefore be strongly structured, and the sparseness of the echo set on display in Figure~\ref{fig:rev}b) is the direct expression of that structuring: the echoes mark the peaks of the field, not a region uniformly above threshold.

Next, we must address the question of what effect extreme events such as cluster 3624 on 10 May 2024 exerts on their environment, and especially the region of marginal growth rates around the cluster. Where the drive is only marginally above threshold the growth rate is small, and the disturbance is carried out of the unstable layer before it accumulates \cite{drexlerNewInsightsNonlocal2002}. The neutral density gradient at high latitudes lies close to the field direction, which gives the marginally unstable modes an upward group velocity that increases as their aspect angle evolves, while the associated parallel electric field is shorted out by field-aligned electron motion \cite{drexlerNewInsightsNonlocal2002}. Efficient conversion of field energy into electron heat \cite{st-mauriceRevisitingBehaviorERegion2021} drives the marginal regions toward this condition. The wide, weakly supercritical surroundings of an extreme event therefore produce no persistent backscatter, while the externally driven cluster persists and is tracked.

Cluster 3624's large envelope and relative sparseness of echoes (Figure~\ref{fig:rev}b) also bears on the reliability of the fastest tracks that we can observe. At 11,240~m/s, the average frame-to-frame displacement is 11.2~km, a small fraction of the envelope's spatial extent, and so the shape overlap on which the assignment depends remains large. 
What is more, Figure~\ref{fig:rev}c) demonstrates that the mean rate at which echoes are detected per unit area within a cluster is essentially independent of cluster speed. In other words, super-fast clusters are not morphologically distinct from the population as a whole. 

\subsection{Proxy measurements for the ionospheric electric field}

As we established in the previous subsection, the $\alpha$-shapes delineate the outer contour of the set of locations (echoes) where the driving field exceeds the FB threshold, near 20~mV/m. The tracked speed is the translation of the echo centre-of-mass within that contour. Three velocities must then be kept apart: \textit{(1)} the phase speed of the waves, near $C_s$ and measured as the radar Doppler shift, \textit{(2)} the $\boldsymbol{E}\times\boldsymbol{B}$-drift of the F-region plasma; and \textit{(3)} the pattern speed of the exceedance (echo) set, which we measure. In Cluster 3624 of Figure~\ref{fig:metop}f), the first (Dopplers) and the third (tracked speeds) \textit{differ by a factor of 45}, with the Doppler speeds staying around 250 m/s, while the echo centre-of-mass inside the contour translates at 11,240 m/s (Figure~\ref{fig:metop}f; see also Figure~\ref{fig:rev}b), and so the tracked object is not the wave population.

That the contour follows the instantaneous field, rather than carrying memory of it, follows from the lifetime of the turbulence. Three-metre irregularities decay within 1 to 2 s once the field falls below threshold \cite{prikryl_doppler_1988,prikryl_evidence_1990}, while tracks can persist beyond 100~s, in which case a single tracked structure encompasses of order one hundred extinctions and re-excitations of the instability. The echoes mark where the field exceeds threshold at the measurement time.

The pattern speed, the tracked motions, equals the $\boldsymbol{E}\times\boldsymbol{B}$-drift \textit{if the field structure is itself frozen into the flow above roughly 150 km altitude}. The same assumption underlies the standard interpretation of four decades worth of coherent-scatter convection measurements \cite{thomasStatisticalPatternsIonospheric2018}. In our data, the tracked speeds match the DMSP F16 ion drifts in the two 2021 conjunctions (Figures \ref{fig:dmsp}, \ref{fig:dmsp2}), and across four years, the speed distribution agrees with Swarm~A cross-track ion drifts to within a factor of two in probability density between 300 and 4000 m/s, and the tracked radar speeds continue without a break through the speed at which the \textit{in-situ} drift moment becomes noise-limited (Figure \ref{fig:stats}).

The foregoing notwithstanding, a translating contour (of FB threshold exceedance, denoted by the displacement of its centroid) does not by itself require displaced plasma. Structures excited \textit{in sequence} at adjacent locations would displace the centroid of an echo distribution. No refinement of the tracking algorithm can exclude this eventuality for any individual cluster, since the same construction may be posited at any scale. The agreement above (Figures \ref{fig:dmsp}f, \ref{fig:dmsp2}d, \ref{fig:stats}a), as well as in the literature \cite{ivarsen_point-cloud_2024,ivarsen_deriving_2024,ivarsen_eastward_2025-1} is what licenses the interpretation that the tracked radar motions are proxy measurements of the ionospheric electric field. As such, splits and mergers reflect changes in the topology of a contour (enclosing FB threshold exceedences) rather than the division of a fluid element. \color{black}

Cluster 3624 is the upper-tail sample of \textsc{icebear}'s wide speed distribution (Figure~\ref{fig:metop}f). Its fitted speed of $11,240\pm660$~m/s implies $|E|\approx560$~mV/m, exceeding the documented STEVE range \citep[100–200 mV/m; ][]{gilliesApparentMotionSTEVE2020,mishinInnerStructureSTEVELinked2023}, on the order of the most extreme reported SAID \cite[around 400 mV/m; ][]{andersonMultisatelliteObservationsRapid2001}. 
Fields of this magnitude have been measured in situ \cite{marklundAuroralPhenomenaRelated1997,marklundObservationsElectricField1998,johanssonIntenseHighaltitudeAuroral2004,labelleElectricFieldStatistics2010} and here acquire the surrounding distribution that places them in statistical context.

\subsection{Utility for space weather models}

\color{black}
Section 4.2 established under which conditions the tracked velocities follow the ionospheric electric field. Here we treat the field's variability, which enters the Joule heating budget through the second moment of $E$.

Here, we note that the auroral electric field governs the local energy budget, through the field's second moment, with the variance carrying the contribution of electric-field structuring to the dissipation \cite{codrescuImportanceEfieldVariability1995}. A global magnetohydrodynamic (MHD) model (a space weather model), which is advanced on a resolved-scale convection pattern, represents this crucial term only down to the model's spatial resolution and update cadence. A sub-grid, sub-cadence  variability is commonly identified as a source of Joule-heating underestimation in models \cite{dengPossibleReasonsUnderestimating2007,cosgroveBiasJouleHeating2011}. The deficit is not alleviated at the $\sim$50–100~km resolution of contemporary global models. In convergence tests the modelled heating rate still rises as latitudinal grid spacing sharpens from 2.5$^\circ$ to 1.25$^\circ$ \cite{dengPossibleReasonsUnderestimating2007}; and the measured variability spectrum retains turbulent power below 45 km \cite{cousinsStatisticalCharacteristicsSmallscale2012}. In global MHD, the electrojet turbulence responsible for part of this variance is not intrinsic to the simulation and must be parameterised in, and can shift the cross-polar-cap potential and conductance by 10–20\% once added \cite{wiltbergerEffectsElectrojetTurbulence2017}. Injecting the observed sub-grid variability can raise the hemispheric Joule-heating rate by a factor $\sim1.5$ \cite{matsuoEffectsHighlatitudeIonospheric2008,matsuoMultiresolutionModelingHighLatitude2021}.

\textsc{icebear} provides some succor in this regard. However, the radar-tracked ensemble is a conditional sample. Echoes exist only where the field exceeds the FB threshold near 20 mV/m, and so the distributions in Figures \ref{fig:dmsp}f), \ref{fig:dmsp2}d), \ref{fig:metop}e) and \ref{fig:noaa}e) describe the field \emph{given} FB threshold-exceedance, and carry no information about the intervening, sub-threshold (stable) regions. What the distributions supply is the conditional mean square, $\langle | E_\perp|^2\rangle_S = \mu_{E,S}^2 + \sigma_{E,S}^2$, with subscript $S$ denoting the unstable regions (or set of echoes), and subscript $E$ denoting the electric field. \color{black}

Aggregated over the four-year radar archive (Figure~\ref{fig:stats}b), \color{black} the conditional dispersion traces geomagnetic activity. A weighted fit gives, \color{black}
\begin{equation} \label{eq:model}
\sigma_{v,S}^{2} = 3.68\times10^{5}\,\bigl(\mathrm{SME}/100\,\mathrm{nT}\bigr)^{0.353}\quad[\mathrm{m^2\,s^{-2}}],
\end{equation} 
\color{black} where $\sigma_{E,S} = |B| \sigma_{v,S}$. This
yields a conditional dispersion \color{black} amplitude $\sigma_{E,S}$ of $\approx$30~mV/m at SME~$=100$~nT rising to $\approx$60~mV/m at SME~$\approx$4000~nT, using typical values of the geomagnetic field \cite{alkenInternationalGeomagneticReference2021}. 

\color{black}
\textsc{icebear} monitors the region of favourable aspect angles within its field of view, whose size and location vary and are not precisely known \cite{ivarsen_turbulence_2024}. Eq.~(\ref{eq:model}) therefore gives the dispersion of the electric field within detected radar aurora \textit{inside} that region, with no information about the exceedance fraction, the unstable fraction of a volume, which is the second quantity the heating budget requires. In this regard, our earlier work characterizes the internal structure of the echo clusters on the scales $75$~km--$3$~km, using pair statistics of the echo point process \cite{ivarsenDistributionSmallScaleIrregularities2023,ivarsenMeasuringSmallscalePlasma2023,ivarsen_turbulence_2024-1}.

The above notwithstanding, we note the shallow exponent in Eq.~(\ref{eq:model}) as reflective of a \textit{weak} activity-dependence of the electric field dispersion inside the active regions, in the conditional (on threshold exceedance) dispersion alone. This distinguishes our parameterizations from those based on unconditional variance \cite{codrescuElectricFieldVariability2000,matsuoEffectsHighlatitudeIonospheric2008}.

As to succor, we note that, \color{black} with $\Sigma_P\sim5$–30~S, the $\approx$560~mV/m field of Cluster~3624 implies a local dissipation rate  $Q=\Sigma_P|E|^2\approx1.5$–10~W/m$^2$, at the upper end of point measurements for major storms \cite{codrescuImportanceEfieldVariability1995,dengPossibleReasonsUnderestimating2007,knippDirectIndirectThermospheric2004}, dissipation that may feed the storm-time budget underestimated by convection-driven models \cite{rosenqvistMagnetosphericEnergyBudget2006} \color{black} (though we note that $1.5$–10~W/m$^2$ is a rate at the exceedances and cannot be integrated over the $\alpha$-shape area, most of which is below the FB threshold; see Figure~\ref{fig:rev}b). \color{black}

Furthermore, \color{black} the FB turbulence serving as the tracer is itself a heating agent, driving anomalous electron heating and possibly raising the Pedersen conductance \cite{dimantMagnetosphereionosphereCouplingRegion2011,oppenheim_kinetic_2013,st-mauriceRevisitingBehaviorERegion2021}. Because this turbulence develops and saturates on timescales shorter than the event itself, a fraction of the imposed field energy is plausibly diverted into \textit{electron thermal energy}, and possibly field-aligned electric fields, before bulk Pedersen (Joule) dissipation is established \cite{schlegelAnomalousHeatingPolar1981,dimantMagnetosphereionosphereCouplingRegion2011}. 
As the observed echo structures are externally driven by precipitation rather than polarization features \cite{ivarsen_point-cloud_2024,ivarsen_turbulence_2024,ivarsen_deriving_2024,ivarsen_characteristic_2025,ivarsen_eastward_2025}, the measured variance enters the budget with positive sign \cite{cosgroveBiasJouleHeating2011}.

\color{black}
In closing, the picture that emerges is of an auroral electric field that is ostensibly \textit{not} a smoothly varying field, but a sparse set of intense, transient structures translating through a background held near the FB instability threshold. It is the occupancy of those structures, rather than their amplitude, that remains to be measured in future investigations.
\color{black}

\section*{Acknowledgements}

This work is supported in part by the European Space Agency’s Living Planet Grant No. 1000012348. We acknowledge the support of the Canadian Space Agency (CSA) [20SUGOICEB], the Canada Foundation for Innovation (CFI) John R. Evans Leaders Fund [32117], the Natural Science and Engineering Research Council (NSERC), the Discovery grants program [RGPIN-2019-19135], the Digital Research Alliance of Canada [RRG-4802], and basic research funding from Korea Astronomy and Space Science Institute
[KASI2026183005]. The authors acknowledge the use of Super\textsc{darn} data. Super\textsc{darn} is a collection of radars funded by national scientific funding agencies of Australia, Canada, China, France, Italy, Japan, Norway, South Africa, United Kingdom and the United States of America.  MFI \& GCH are grateful to P. Erickson and F. Lind for stimulating discussions. Anthropic's Claude Opus 4.6 \& 4.7, and Google Scholar Labs, were used to assist research, and Anthropic's Claude Opus 4.7 was used to assist coding in \textsc{matlab}. The first author verified generated output and takes responsibility for it.

\section*{Open Research}

\textsc{icebear} Level 1 data is available at \url{http://icebear.usask.ca/icebeardata.php?page=1}; Level 2 data for 2020, 2021 is available with DOI \url{https://doi.org/10.5281/zenodo.7509022} \cite{huyghebaertICEBEARAlldigitalBistatic2019,lozinskyICEBEAR3DLowElevation2022}. Super\textsc{mag} data (as well as Rabbit Lake magnetometer data) can be accessed at \url{https://supermag.jhuapl.edu/mag/} \cite{gjerloevSuperMAGDataProcessing2012}. Super\textsc{darn} data was processed using RST with the FITACF 3 algorithm, order/degree 6, Ref.~\cite{thomasStatisticalPatternsIonospheric2018}-empirical model, downloadable using Globus  (\url{https://github.com/SuperDARNCanada/globus}). \textsc{metop} and \textsc{noaa} data is available from \textsc{noaa} (\url{https://www.ngdc.noaa.gov/stp/ satellite/poes/}). Data from DMSP's SSJ and SSIES instruments can be accessed through Madrigal (\url{http://cedar.openmadrigal.org/}). Code that implements the full tracking methodology is available at Zenodo \cite{ivarsenPredictiveTrackingAlgorithm2026}.


\end{document}